**Aleksandr CARIOW[1]**, Galina CARIOWA[1] and Marina CHICHEVA[2,3]
[1]WEST POMERANIAN UNIVERSITY OF TECHNOLOGY, SZCZECIN, Żołnierska St. 49, 71-210, Szczecin Poland
[2]SAMARA UNIVERSITY Moskovskoye sh., 34, Samara, Russia, 443086
[3]IPSI RAS – Branch of the FSRC "Crystallography and Photonics" RAS, 151, Samara, Russia, 443001

# Hardware-Efficient Schemes of Quaternion Multiplying Units for 2D Discrete Quaternion Fourier Transform Processors

**Abstract**

In this paper, we offer and discuss three efficient structural solutions for the hardware-oriented implementation of discrete quaternion Fourier transform basic operations with reduced implementation complexities. The first solution - a scheme for calculating *sq* product, the second solution – a scheme for calculating *qt* product, and the third solution – a scheme for calculating *sqt* product, where *s* is a so-called *i* -quaternion, *t* is an *j* - quaternion, and *q* – is an usual quaternion. The direct multiplication of two usual quaternions requires 16 real multiplications (or two-operand multipliers in the case of fully parallel hardware implementation) and 12 real additions (or binary adders). At the same time, our solutions allow to design the computation units, which consume only 6 multipliers plus 6 two input adders for implementation of *sq* or *qt* basic operations and 9 binary multipliers plus 6 two-input adders and 4 four-input adders for implementation of *sqt* basic operation.

## 1. Introduction

Two dimension discrete Fourier transform (2D-DFT) have been widely used in image processing ever since the discovery of Fast Fourier transform (FFT) which made the computation of DFT feasible using a computer [1]. However, if we want to apply the classical 2D-FFT to color images, we must perform three separate 2D-FFTs. This is because every color image pixel has three values associated with it: the red, green, and blue components.

Until recently, it was not offered any discrete transform, which would perceive the each pixel of the color image as a whole. Sangwine and Ell defined a new transform, called the discrete quaternion Fourier transform (DQFT) which allows to process simultaneously of all color components of the image [2-4]. The idea of DQFT is based on representation of color image pixels via quaternions – four-dimensional hypercomplex numbers discovered by Hamilton in 1843. Today there are several ways to calculate the 2D-DQFT [5-13]. Another way for calculating the QDFT has been reported in [14]. While it produces the same result as the other approaches, it is more efficient because leads to reducing the computation complexity. During the DQFT implementation using the method proposed in [14], it is necessary to perform three types of the quaternion multiplication operation, namely: left-sided quaternion multiplication, right-sided quaternion multiplication and two-sided quaternion multiplication. What is more, in all three cases only one quaternion is a usual quaternion, and the rest quaternions are constant quaternions, i.e. quaternions, which coefficients are real constants. Next we propose hardware-effective schemes to implement these operations.

## 2. Statement of the problem

The typical operations of the related two dimensional forward and inverse discrete quaternion Fourier transform are [14]:

- left-sided quaternion multiplication $sq$,
- right-sided quaternion multiplication $qt$ and
- two-sided quaternion multiplication $sqt$,

Where $q = q_0 + q_1 i + q_2 j + q_3 k$ is a conventional quaternion, with three imaginary units $ij = k$, $ji = -k$, $i^2 = j^2 = k^2 = ijk = -1$ and four real variables $\{q_n\}$, $n = 0,1,2,3$; $s = \alpha + \beta i + 0 j + 0k$, and $t = \gamma + 0i + \delta j + 0k$ are so-called $i$ – quaternion and $j$ – quaternion respectively [14], and $\alpha, \beta, \gamma, \delta$ - are real constants.

During synthesis of the discussed schemes we use the fact that multiplication of two quaternions may be represented as vector-matrix product [15, 16]. The matrix that participates in the product calculating has unique structural properties that allow performing its advantageous factorization [17]. Namely this factorization leads to significant reducing of the computational complexity of quaternion multiplication. Furthermore, since $s$ and $t$ are truncated quaternions and are in fact complex numbers, which located in the different domains of complex space, the corresponding matrices are sparse. This leads to additional effect in minimization of the computational complexity. Finally, an additional effect can be achieved using the fact that the numbers $\alpha$, $\beta$, $\gamma$ and $\delta$ are real constants and their products can be calculated and stored in memory in advance.

## 3. The schemes

Let $\mathbf{X}_{4\times 1} = [q_0, q_1, q_2, q_3]^T$ - be a column vector, that contains the all coefficients of quaternion $q$, and $\mathbf{Y}_{4\times 1}^{(1)} = [y_0^{(1)}, y_1^{(1)}, y_2^{(1)}, y_3^{(1)}]^T$ - be a column vector containing the elements of $sq$ product.

The first scheme (for implementation $sq$ - kernel) can be written with the help of following matrix-vector calculating procedure:

$$\mathbf{Y}_4^{(1)} = \mathbf{P}_4^{(1)} \mathbf{A}_{4\times 6}^{(1)} \mathbf{D}_6^{(1)} \mathbf{A}_{6\times 4}^{(1)} \mathbf{X}_{4\times 1} \quad (1)$$

where

$$\mathbf{A}_{6\times 4}^{(1)} = \begin{bmatrix} 1 & -1 & & \\ 1 & 0 & \mathbf{0}_{3\times 2} & \\ 0 & 1 & & \\ \hline & & 1 & -1 \\ \mathbf{0}_{3\times 2} & & 1 & 0 \\ & & 0 & 1 \end{bmatrix},$$

$$\mathbf{D}_6^{(1)} = diag(\alpha, d_1, d_2, d_1, d_2, \alpha), \quad d_1 = \alpha + \beta, \quad d_2 = \alpha - \beta,$$

$$\mathbf{A}_{4\times 6}^{(1)} = \begin{bmatrix} -1 & 1 & 0 & & \\ 1 & 0 & 1 & \mathbf{0}_{2\times 3} \\ \hline & & & -1 & 1 & 0 \\ & \mathbf{0}_{2\times 3} & & 1 & 0 & 1 \end{bmatrix}, \quad \mathbf{P}_4^{(1)} = \begin{bmatrix} 0 & 1 & & \\ 1 & 0 & \mathbf{0}_2 & \\ \hline & & 0 & 1 \\ & \mathbf{0}_2 & 1 & 0 \end{bmatrix}.$$

Fig. 1 shows a data flow diagram of the proposed scheme for realization $sq$ product kernel/ In this paper, data flow diagrams are oriented from left to right. Straight lines in the figures denote the operations of data transfer. The circles in these figures show the operation of multiplication by a real number inscribed inside a circle. Points where lines converge denote summation a dotted lines indicate the sign-change operations. We use the usual lines without arrows on purpose, so as not to clutter the picture.

Fig. 1. The data flow diagram of proposed scheme for implementation *sq* product kernel

Let $\mathbf{Y}_{4\times 1}^{(2)} = [y_0^{(2)}, y_1^{(2)}, y_2^{(2)}, y_3^{(2)}]^T$ - be a column vector containing the elements of *qt* product.

Then the second scheme (for implementation *qt* - kernel) can be written with the help of following matrix-vector calculating procedure:

$$\mathbf{Y}_4^{(2)} = \mathbf{P}_4^{(2)} \mathbf{A}_{4\times 6}^{(2)} \mathbf{D}_6^{(2)} \mathbf{A}_{6\times 4}^{(2)} \mathbf{P}_4^{(2)} \mathbf{X}_{4\times 1} \quad (2)$$

where

$$\mathbf{P}_4^{(2)} = \begin{bmatrix} & & & 1 \\ 1 & & & \\ & & 1 & \\ & 1 & & \end{bmatrix}, \quad \mathbf{A}_{6\times 4}^{(2)} = \begin{bmatrix} 1 & 0 & & \\ 1 & 0 & \mathbf{0}_{3\times 2} & \\ -1 & 1 & & \\ & & 1 & -1 \\ \mathbf{0}_{3\times 2} & & 0 & 1 \\ & & 0 & 1 \end{bmatrix},$$

$$\mathbf{A}_{4\times 6}^{(2)} = \begin{bmatrix} 1 & 0 & 1 & & & \\ 0 & 1 & 1 & \mathbf{0}_{2\times 3} & & \\ & \mathbf{0}_{2\times 3} & & 1 & 0 & 1 \\ & & & 0 & 1 & 1 \end{bmatrix}, \quad \mathbf{P}_4^{(3)} = \begin{bmatrix} 1 & & & \\ & & 1 & \\ & 1 & & \\ & & & 1 \end{bmatrix}.$$

$\mathbf{D}_6^{(1)} = diag(\alpha, d_1, d_2, d_1, d_2, \alpha)$, $g_1 = \gamma - \delta$, $g_2 = \gamma + \delta$.

Fig. 2 shows a data flow diagram of the proposed scheme for implementation *qt* product kernel.

Fig. 2. The data flow diagram of proposed scheme for implementation *qt* product kernel

Let $\mathbf{Y}_{4\times 1}^{(3)} = [y_0^{(3)}, y_1^{(3)}, y_2^{(3)}, y_3^{(3)}]^T$ - be a column vector containing the elements of *sqt* product. Then the third scheme (for implementation *sqt* - kernel) can be written with the help of following matrix-vector calculating procedure:

$$\mathbf{Y}_{4\times 1}^{(3)} = \mathbf{A}_{4\times 9} \mathbf{D}_9 \mathbf{W}_{9\times 7} \mathbf{W}_7 \mathbf{A}_{7\times 4} \mathbf{P}_4 \mathbf{X}_{4\times 1} \quad (3)$$

$$\mathbf{A}_{7\times 4} = \begin{bmatrix} 1 & & & \\ & 1 & & \\ & & 1 & \\ & & 1 & -1 \\ & & & 1 \\ & 1 & & \\ & & & 1 \end{bmatrix}, \quad \mathbf{W}_7 = \begin{bmatrix} \mathbf{H}_2 & \mathbf{0}_{2\times 3} & \mathbf{0}_2 \\ \mathbf{0}_{3\times 2} & \mathbf{I}_3 & \mathbf{0}_{3\times 2} \\ \mathbf{0}_2 & \mathbf{0}_{2\times 3} & \breve{\mathbf{H}}_2 \end{bmatrix} =$$

$$= \begin{bmatrix} 1 & 1 & & & & & \\ 1 & -1 & & & & & \\ & & 1 & & & & \\ & & & 1 & & & \\ & & & & 1 & & \\ & & & & & 1 & -1 \\ & & & & & 1 & 1 \end{bmatrix},$$

$$\mathbf{W}_{9\times 7} = \begin{bmatrix} 1 & & & & & & \\ & 1 & & & & & \\ & 1 & & & & & -1 \\ & & 1 & & & & \\ & & & 1 & & & \\ & & & & 1 & & \\ 1 & & & & & -1 & \\ & & & & & 1 & \\ & & & & & & 1 \end{bmatrix}, \quad \mathbf{P}_4 = \begin{bmatrix} 1 & & & \\ & & 1 & \\ & 1 & & \\ & & & 1 \end{bmatrix}$$

$$\mathbf{D}_9 = diag(p_6, p_5, p_2, p_4, p_1, p_4, p_3, p_6, p_5)$$

where

$p_1 = (\alpha - \beta)$, $p_2 = \alpha\delta$, $p_3 = \beta\delta$, $p_4 = (\alpha-\beta)(\gamma-\delta)$, $p_5 = \alpha(\gamma-\delta)$, $p_6 = \alpha(\gamma-\delta)$.

Fig. 3 shows a data flow diagram of the proposed scheme for implementation *qt* product kernel. The rectangles indicate the operations of multiplication by the matrices $\mathbf{H}_2$ and $\breve{\mathbf{H}}_2$.

Fig. 3. The data flow diagram of proposed scheme for implementation *sqt* product kernel

3## 4. Conclusion

The article presents three new hardware-efficient schemes for the execution *sq* - product, *qt* - product and *sqt* - product kernels with reduced computational complexities. To reduce the hardware complexity (number of embedded adders and multipliers), we exploit the specific structural properties of the matrix-vector products that represent mentioned basic operations. So, the fully parallel implementation of *sq* - product and *qt* - product kernels require only 6 multipliers by real numbers, and 6 adders. In turn, a fully parallel implementation of *sqt* - product kernel requires only 9 binary multipliers, 6 two-input adders and 4 four-input adders.

Reducing the number of multiplications is especially important in the design of specialized VLSI on-board DSP processors because minimizing the number of necessary multipliers also reduces the power dissipation and lowers the cost implementation of the entire system being implemented. This is because a hardware multiplier is more complicated unit than an adder and occupies much more chip area than the adder. (It is proved that the hardware complexity of an embedded multiplier grows quadratically with operand size, while the hardware complexity of an binary adder increases linearly with operand size). Even if the VLSI chip already contains embedded multipliers, their number is always limited. This means that if the implemented scheme has a large number of multiplications, the projected processor may not always fit into the chip and the problem of minimizing the number of multipliers remains relevant.

This problem becomes extremely challenging for applications requiring real-time processing at high throughput especially in digital signal and image processing. Hence, for meeting the high requirements to throughput and power-consumption constraints of real-time image processing systems, developing hardware-efficient schemes to implement them on the base of application specific integrated circuits (ASICs) or field-programmable gate arrays (FPGAs) is of paramount importance.

**Prof., DSc., PhD Aleksandr CARIOW**

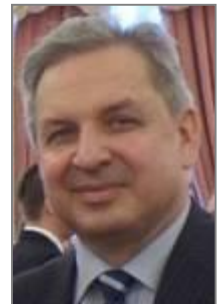

He received the Candidate of Sciences (PhD) and Doctor of Sciences degree (DSc or Habilitation) in Computer Sciences from LITMO of St. Petersburg, Russia in 1984 and 2001, respectively. In September 1999, he joined the faculty of Computer Sciences and Information Technology at the West Pomeranian University of Technology, Szczecin, Poland, where he is currently a professor and chair of the Department of Computer Architectures and Telecommunications. His research interests include digital signal processing algorithms, VLSI architectures, and data processing parallelization.

*e-mail: acariow@wi.zut.edu.pl*

**PhD Galina CARIOWA**

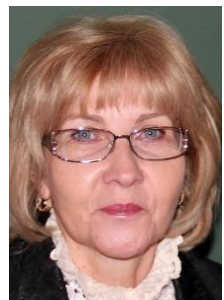

She received the MSc degree in Mathematics from Moldavian State University, Chişinău in 1976 and PhD degree in computer science from West Pomeranian University of Technology, Szczecin, Poland in 2007. She is currently working as an assistant professor of the Department of Multimedia Systems. She is also an Associate-Editor of World Research Journal of Transactions on Algorithms. Her scientific interests include numerical linear algebra and digital signal processing algorithms, VLSI architectures, and data processing parallelization.

*e-mail: gcariowa@wi.zut.edu.pl*

**PhD Marina CHICHEVA**

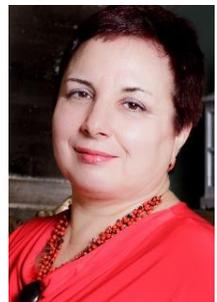

She received the MSc and PhD degrees in Computer Science from the Kuibyshev Aviation Institute (now Samara State Aerospace University) in 1987 and in 1998 respectively. She is currently working as assistant professor in the Samara University and as a senior researcher in the Image processing systems institute – branch of the Federal Scientific Research Centre "Crystallography and photonics" of Russian Academy of Sciences. Her scientific interests include image processing, data compression, and fast algorithms of discrete transforms. She is also a Member of the Russian Association of Pattern Recognition and Image Analysis.

*e-mail: mchi@geosamara.ru*